\documentclass[conference, comsoc]{IEEEtran}
\IEEEoverridecommandlockouts
\usepackage[left=0.625in,right=0.625in,top=0.75in,bottom=1in,columnsep=0.21in]{geometry}

\usepackage{balance}
\usepackage{caption}
\usepackage{subcaption}
\usepackage{multirow}
\usepackage{array}
\usepackage{balance}
\usepackage{cite}
\usepackage{amsmath,amssymb,amsfonts}
\usepackage{graphicx}
\usepackage{booktabs}
\usepackage{textcomp}
\usepackage{bm}
\usepackage{nicefrac}
\usepackage{mathtools}
\usepackage{comment}
\usepackage{pgfplots}
\usepackage{pgfkeys,pgfmath,pgfcore}
\pgfplotsset{compat=1.18}
\pgfkeys{/pgf/number format/.cd,fixed,precision=2}

\usepackage[ruled, vlined]{algorithm2e}
\usepackage{multirow}
\usepackage{xcolor} 
\usepackage{colortbl}
\usepackage{threeparttable}
\usepackage{dblfloatfix} 
\usepackage{acro}
\definecolor{verylightgray}{rgb}{0.9,0.9,0.9}
\usepackage{tikz}
\usetikzlibrary{arrows.meta,bending,calc}
\usepackage{algorithmic}
\usepackage{booktabs}

\newif\ifshowtodos
\showtodostrue

\DeclareAcronym{wfm}{
  short = WFM,
  long  = Wireless Foundation Model
}

\DeclareAcronym{lora}{
  short = LoRA,
  long = Low-Rank Adaptation
}

\DeclareAcronym{jepa}{
  short = JEPA,
  long = Joint-Embedding Predictive Architecture
}

\DeclareAcronym{ema}{
    short = EMA,
    long = Exponential Moving Average
}

\DeclareAcronym{sdr}{
    short = SDRs,
    long = Software-Defined Radios
}

\DeclareAcronym{vit}{
    short = ViT,
    long = Vision Transformer
}

\begin{document}

\title{LatentWave: JEPA Pretraining for \\Wireless Foundation Models}

\author{\IEEEauthorblockN{Ahmed Mohamed, Ahmed Aboulfotouh, and
Hatem Abou-Zeid}

\thanks{This work was supported in part by Alberta Innovates and in part by the Natural Sciences and Engineering Research Council of Canada (NSERC) under Grant RGPIN-2021-04050.}
\IEEEauthorblockA{{Department of Electrical and Software Engineering}, 
{University of Calgary}, Canada}}

\maketitle

\begin{abstract} Wireless foundation models have emerged as a promising alternative to building separate models for each wireless task. However, existing approaches rely on masked input reconstruction, which can bias representations toward low-level signal details. In this paper, we propose LatentWave, a wireless foundation model pretrained using a Joint-Embedding Predictive Architecture (JEPA) on diverse wireless spectrograms and channel state information (CSI). By predicting masked regions in latent space, LatentWave learns representations that are more transferable out of the box across diverse downstream tasks. The proposed architecture employs per-channel patch embeddings with stochastic channel sampling during pretraining, allowing it to process variable antenna counts and improving usability across heterogeneous wireless configurations. We evaluate LatentWave on four downstream tasks: RF signal classification, 5G NR positioning, beam prediction, and LoS/NLoS classification, comparing against a masked-modeling baseline (WavesFM) pretrained on the same data. Additionally, we show that the masking geometry introduces a task-dependent inductive bias: frequency masking strongly favors channel-related tasks such as positioning and beam prediction, while region masking better preserves discriminability for signal classification.

\end{abstract}

\section{Introduction}

 Conventional wireless AI systems require building a separate model for each task and environment, creating a growing collection of specialized pipelines~\cite{wai} that are expensive to develop, deploy, and maintain. 
 Foundation models and self-supervised learning~\cite{wfm} offer a compelling alternative: by learning general patterns from diverse, unlabeled data, a single pretrained model can be rapidly adapted to new downstream tasks, replacing many specialized systems with one general and adaptable architecture.
In the context of network softwarization, such an architecture can be viewed as a reusable software intelligence layer that maps heterogeneous wireless measurements into task-agnostic representations. Rather than deploying separate AI models for each wireless service, the \ac{wfm} provides a shared substrate that can support multiple downstream functions such as sensing, localization, and beam management.

Recent advances have been made toward this paradigm in the wireless domain, including the works in~\cite{wavesfm, wirelessgpt,iqfm,mmwfm,csi2vec}. The dominant pretraining strategy for \acp{wfm} has been masked modeling with pixel-space reconstruction. However, this approach has an important limitation: by requiring the model to reconstruct raw input values, the model is forced to dedicate representational capacity to low-level spectral details, noise textures, and fine-grained amplitude variations that may carry little discriminative value for downstream tasks. 
Results indicate that \acp{wfm} pretrained with pixel-space masking and reconstruction require either fine-tuning of several last layers or parameter-efficient adaptation to achieve competitive performance on tasks that differ substantially from the pretraining distributions.
%
This suggests that the frozen representations alone do not fully capture the high-level semantic structure needed for diverse wireless tasks.

The \ac{jepa} framework~\cite{jepa} offers an alternative: instead of reconstructing missing pixels, \ac{jepa} trains a model to predict the representations of \emph{masked regions} in a learned latent space. Because the prediction target is itself a learned abstraction, the model is encouraged to capture higher-level semantic features while discarding irrelevant low-level variation. However, the effectiveness of such an approach has not been explored for wireless modalities and tasks that leverage CSI and spectrograms. 

\begin{figure*}
    \centering
    \includegraphics[width=\linewidth]{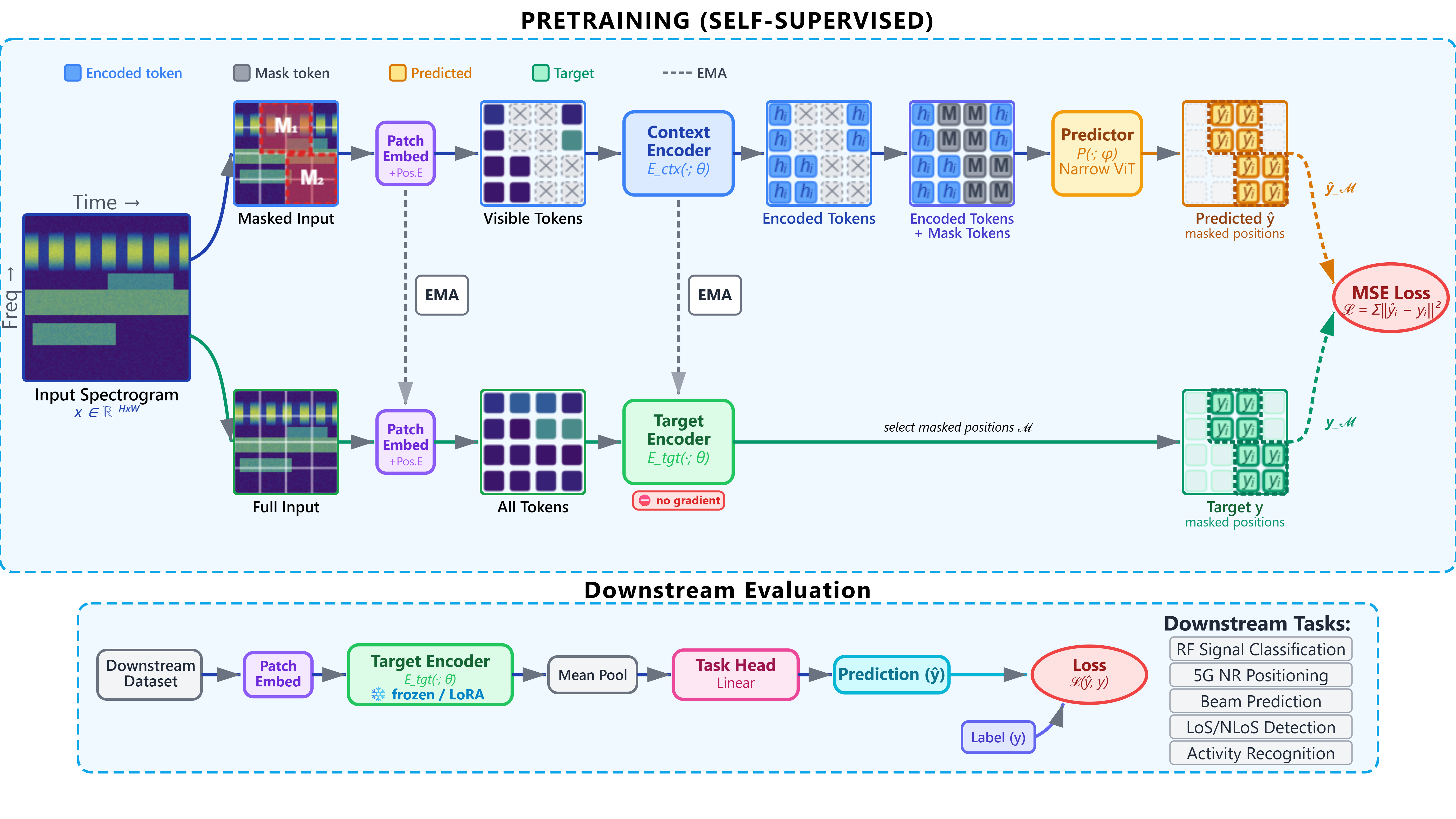}
    \caption{LatentWave architecture. Pretraining (top): a JEPA-based self-supervised approach predicts latent representations of masked spectrogram patches. Downstream evaluation (bottom): the pretrained encoder is adapted to downstream wireless tasks.}
    \label{fig:LatentWave}
\end{figure*}

In this paper, we present \textbf{LatentWave}, a \ac{wfm} pretrained with \ac{jepa} on a diverse dataset of wireless spectrograms and CSI. We evaluate the learned representations on four downstream tasks spanning communication, sensing, and positioning. Our main contributions are as follows:

\begin{itemize} 
\item We propose \textbf{LatentWave}, the first \ac{jepa}-based \ac{wfm} designed to support diverse downstream wireless tasks involving both spectrogram and CSI inputs. The architecture employs a per-channel patch embedding that allows processing of variable antenna counts, and a stochastic channel sampling strategy during pretraining that exposes the model to varying antenna configurations, enabling generalization across heterogeneous wireless setups.
\item We conduct a systematic comparison of four masking strategies: region, frequency, time, and random with varying numbers of masks. We show that the masking geometry introduces a task-dependent inductive bias: frequency masking favors channel-related tasks such as positioning and beam prediction, while region and time masking better preserve temporal discriminability needed for signal classification. No single strategy dominates all tasks. 
\end{itemize}

\section{Related Work}

Several recent works have explored the development of wireless foundation models.
WavesFM~\cite{wavesfm} demonstrated that masked modeling can learn broadly transferable representations from image-like wireless modalities, including spectrograms and CSI, jointly supporting sensing, localization, and communication tasks.
Other notable efforts have proposed foundation models for CSI feature extraction, using masked modeling as the pretraining objective with evaluation on line-of-sight/non-line-of-sight classification and beam prediction~\cite{wirelessgpt}. Contrastive learning has also been explored as an alternative self-supervised strategy: CSI2Vec~\cite{csi2vec} employs a triplet-based contrastive objective and validates its representations on positioning and channel charting, while IQFM~\cite{iqfm} learns general-purpose features from raw in-phase and quadrature~(IQ) signals for modulation classification, RF device fingerprinting, and angle-of-arrival estimation, demonstrating strong sample efficiency in few-shot settings. Multi-modal approaches~\cite{mmwfm} that jointly handle spectrograms, CSI, and IQ signals have also been investigated. 

More recently, WirelessJEPA~\cite{wirelessjepa} applied CNNs within a latent prediction framework to learn representations of IQ signals, outperforming contrastive approaches on multiple downstream tasks. However, this work is limited to IQ signals in both pretraining and downstream.
More recently, the work in~\cite{naoumi2026homomorphic} proposed a JEPA-based approach to model the temporal evolution of CSI.
Their results show strong long-horizon prediction, suggesting that JEPA-based learning can capture meaningful temporal wireless dynamics in the form of CSI embeddings. While this work is valuable for temporal CSI representation learning and prediction, it does not focus on learning general-purpose features for wireless foundation models that would enable
 a broad and diverse set of downstream wireless tasks.


\section{LatentWave Methodology}
This section presents the LatentWave methodology, including the architecture and masking strategies, training algorithms, and datasets for pretraining and downstream tasks.

\subsection{Proposed Pretraining Framework and Architecture}

Our foundation model is pretrained on a diverse, unlabeled dataset of wireless signal spectrograms and CSI using the \ac{jepa} framework~\cite{jepa}. The core principle is to learn representations by predicting in a learned latent space rather than reconstructing raw input data, as shown in Fig.~\ref{fig:LatentWave}. 
\ac{jepa} is a self-supervised framework that operates on masked inputs through three interacting components: a \emph{context encoder}, a \emph{target encoder}, and a \emph{predictor}. Given an input, a subset of its content is masked. The context encoder processes only the visible (unmasked) portions and maps them to latent representations. The target encoder, which shares the same architecture, processes the complete input and produces latent targets. The predictor then forecasts the target encoder's representations of the masked regions using only the context encoder's output. Predictions occur in the latent space rather than in the raw input space. This process is illustrated in detail in~\cite{jepa}.

We now describe our instantiation of this framework for wireless data. Let \(x \in \mathbb{R}^{C \times H \times W}\) denote input signals with \(C\) antennas, \(H\) frequency bins and \(W\) time steps. Each antenna channel $x^{(c)} \in \mathbb{R}^{H \times W}$, $c = 1, \ldots, C$, is independently divided into a grid of $\frac{H}{p_h} \times \frac{W}{p_w}$ non-overlapping patches of size $p_h \times p_w$. A shared linear projection maps each single-channel patch to a $D$-dimensional embedding. This yields a total of $N = C \times \frac{H}{p_h} \times \frac{W}{p_w}$ patch tokens $\mathbf{z} = [z_1, z_2, \ldots, z_N] \in \mathbb{R}^{N \times D}$. Because the projection operates on individual channels rather than across all $C$ channels simultaneously, the patch embedding layer is decoupled from the number of input antennas, allowing the same model to ingest inputs with an arbitrary number of channels without architectural modification. Sinusoidal positional embeddings are added to each token to encode its spatial position within the signal grid.

To improve robustness against varying channel configurations, we employ a stochastic channel sampling strategy during pretraining. For each training sample with $C$ available channels, we uniformly draw a random integer $C' \sim \mathcal{U}\{1, C\}$ and retain a randomly selected subset of $C'$ channels, discarding the rest. The effective token sequence length thus varies as $N = C' \times \frac{H}{p_h} \times \frac{W}{p_w}$ across iterations, exposing the model to diverse antenna configurations. 

During pretraining, we adopt a multi-block masking strategy in which several contiguous blocks spanning meaningful time-frequency-antenna regions are masked (rather than random individual patches). This encourages the model to learn high-level semantic structure in order to predict the missing regions, as local interpolation alone is insufficient. The masked patch indices form the set \(\mathcal{M} \subset \{1, \ldots, N\}\), and the complementary visible set is defined as \(\mathcal{V} = \{1, \ldots, N\} \setminus \mathcal{M}\). Full details of the masking strategy and its hyperparameters are provided in Section~\ref{subsec:masking}.

\textbf{Context Encoder.} The context encoder \(E_{\mathrm{cxt}}(\,\cdot\,; \theta)\) is a standard \ac{vit}~\cite{vits}. It receives only the visible patch tokens \(\{z_i\}_{i \in \mathcal{V}}\) and produces their latent representations:
\begin{equation}
    h = E_{\mathrm{cxt}}\!\bigl(\{z_i\}_{i \in \mathcal{V}};\, \theta\bigr) \in \mathbb{R}^{|\mathcal{V}| \times D}.
\end{equation}
By processing only the unmasked subset, the context encoder is encouraged to build contextual representations from incomplete observations.

\textbf{Target Encoder.} The target encoder \(E_{\mathrm{tgt}}(\,\cdot\,; \bar{\theta})\) shares the same \ac{vit} architecture as the context encoder but processes the \emph{complete} sequence of patch tokens (over the $C'$ sampled channels) to produce the latent targets:
\begin{equation}
    y = E_{\mathrm{tgt}}\!\bigl(\{z_i\}_{i=1}^{N};\, \bar{\theta}\bigr) \in \mathbb{R}^{N \times D}.
\end{equation}
No gradients flow through the target encoder. Instead, its parameters are updated as an \ac{ema} of the context encoder weights:
\begin{equation}
    \bar{\theta}_{t} = \tau\, \bar{\theta}_{t-1} + (1 - \tau)\, \theta_{t},
\end{equation}
where \(\tau \in (0,1)\) is the momentum coefficient, \(\theta\) are the context encoder parameters, and $t$ is the update step. This asymmetric update rule ensures the target encoder evolves smoothly to provide stable training targets, preventing representation collapse.

\textbf{Predictor.} The predictor \(P(\,\cdot\,; \phi)\) is a smaller, narrower \ac{vit}. It takes the context encoder output \(h\) and forecasts the latent representations corresponding to the masked positions:
\begin{equation}
    \hat{y} = P\!\bigl(h;\, \phi\bigr) \in \mathbb{R}^{|\mathcal{M}| \times D}.
\end{equation}
where \(\phi\) denotes the predictor parameters. The predictor's limited capacity is deliberate: it forces the context encoder to produce maximally informative representations rather than offloading the prediction task to the predictor.

The model is trained end-to-end to minimize the mean squared error between the predicted and target representations over the masked positions:
\begin{equation}
    \mathcal{L} = \frac{1}{|\mathcal{M}|} \sum_{i \in \mathcal{M}} \bigl\| \hat{y}_i - y_i \bigr\|_2^2,
\end{equation}

with gradients flowing only through the context encoder parameters \(\theta\) and the predictor parameters \(\phi\). The complete pretraining procedure is summarized in Algorithm~\ref{alg:pretraining}. 

\textbf{LatentWave Foundation Model.} After pretraining, the target encoder is extracted as the foundation model for downstream tasks. Because it processed complete spectrograms throughout pretraining without any masking, its representations are directly aligned with the full inputs encountered during fine-tuning, unlike the context encoder, which was trained exclusively on partial observations.

The context and target encoders are \ac{vit} models with 8 transformer layers, 8 attention heads, embedding dimension $D{=}256$, and patch size $16{\times}16$, totaling approximately $6.4$M parameters. The predictor is a smaller \ac{vit} with 4 layers, 4 attention heads, embedding dimension 128, and approximately 880K parameters. The full pretraining hyperparameters are shown in Table~\ref{tab:hyperparams}.

\begin{algorithm}[htbp]
\caption{LatentWave JEPA Pretraining}
\label{alg:pretraining}
\begin{algorithmic}[1]
\REQUIRE Dataset $\mathcal{D}$, epochs $E$, total steps $T$, EMA\\bounds $[\tau_{\min}, \tau_{\max}]$
\vspace{0.5em}
\STATE $\triangleright$ \textit{Initialization}
\STATE Initialize context encoder $E_{cxt}(\theta)$ and predictor $P(\phi)$
\STATE $E_{tgt}(\bar{\theta}) \gets \text{copy}(E_{cxt}(\theta))$ \COMMENT{Initialize target encoder}
\STATE Disable gradient tracking for $E_{tgt}$ \COMMENT{EMA-updated.}
\STATE $t \gets 0$ \COMMENT{Global step counter}

\vspace{0.5em}
\FOR{$e = 1$ \textbf{to} $E$}
    \FOR{each mini-batch $x \in \mathcal{D}$}
        \STATE $t \gets t + 1$
        \vspace{0.5em}
        \STATE $\triangleright$ \textbf{Multi-block masking}
        \STATE Sample $N_m$ target masks $\{M_t^{(i)}\}_{i=1}^{N_m}$
        \STATE $M_c \gets \text{SampleContext}(x,\; \{M_t^{(i)}\})$ \COMMENT{Context\\mask excluding target regions}
        \STATE $x_{vis} \gets \text{Apply}(x,\; M_c)$ \COMMENT{Extract visible patches\\via context mask}
        \vspace{0.5em}
        \STATE $\triangleright$ \textbf{Context encoding and prediction}
        \STATE $h \gets E_{cxt}(x_{vis};\; \theta)$ \COMMENT{Encode visible patches}
        \STATE $\hat{y} \gets P(h,\; \text{pos}(\{M_t^{(i)}\});\; \phi)$ \COMMENT{Predict targets\\with positional info}
        \vspace{0.5em}
        \STATE $\triangleright$ \textbf{Target representation}
        \STATE $y_{\text{full}} \gets E_{tgt}(x;\; \bar{\theta})$ \COMMENT{Target encoder on full input}
        \STATE $y_{norm} \gets \text{LayerNorm}(y_{\text{full}})$ \COMMENT{Normalize}
        \STATE $y^{(i)} \gets \text{Extract}(y_{norm},\ M_t^{(i)})$ for $i = 1, \ldots, N_m$ \COMMENT{Extract target representations}
        \vspace{0.5em}
        \STATE $\triangleright$ \textbf{Loss computation and optimization}
        \STATE $\mathcal{L} \gets \frac{1}{N_m}\sum_{i=1}^{N_m} \text{MSE}\!\left(\hat{y}^{(i)},\; y^{(i)}\right)$ \COMMENT{Average loss\\over masks}
        \STATE Update $\theta, \phi$ via gradient descent on $\mathcal{L}$
        \vspace{0.5em}
        \STATE $\triangleright$ \textbf{EMA and schedule updates}
        \STATE $\tau \gets \tau_{\min} + \frac{t}{T}\left(\tau_{\max} - \tau_{\min}\right)$ \COMMENT{Linear EMA\\schedule}
        \STATE $\bar{\theta} \gets \tau\,\bar{\theta} + (1 - \tau)\,\theta$ \COMMENT{EMA update of target\\encoder}
        \STATE Update learning rate \COMMENT{Cosine schedule}
    \ENDFOR
\ENDFOR
\end{algorithmic}
\end{algorithm}

\begin{table}[t]
\caption{Pretraining Hyperparameters}
\label{tab:hyperparams}
\centering
\begin{tabular}{lc}
\toprule
\textbf{Hyperparameter} & \textbf{Value} \\
\midrule
\multicolumn{2}{l}{\textit{Encoder Architecture}} \\
Input size (\(H \times W\)) & $224 \times 224$ \\
Patch size (\(p_h \times p_w\)) & $16 \times 16$ \\
Embedding dimension ($D$) & 256 \\
Transformer layers & 8 \\
Attention heads & 8 \\
Parameters & $\approx$ 6.4M \\
\midrule
\multicolumn{2}{l}{\textit{Predictor Architecture}} \\
Embedding dimension & 128 \\
Transformer layers & 4 \\
Attention heads & 4 \\
Parameters & $\approx$ 880K \\
\midrule
\multicolumn{2}{l}{\textit{Masking Strategy}} \\
Number of target masks ($N_m$) & $[2, 6]$ \\
Target mask scale ($[s_{\min},\; s_{\max}]$) & $[0.15,\; 0.2]$ \\
Aspect ratio ($[a_{\min},\; a_{\max}]$) & $[0.75,\; 1.5]$ \\
Context mask scale ($[s_{c, \min},\; s_{c, \max}]$) & $[0.85,\; 1.0]$ \\
Minimum visible patches & 10 \\
\midrule
\multicolumn{2}{l}{\textit{Optimization}} \\
Epochs & 240 \\
Warmup epochs & 12 \\
Batch size & 256 \\
Initial learning rate & $1 \times 10^{-6}$ \\
Reference learning rate & $1 \times 10^{-3}$ \\
Final learning rate & $1 \times 10^{-6}$ \\
Weight decay & 0.04 \\
EMA momentum ($[\tau_{\min},\; \tau_{\max}]$) & $[0.996,\; 1.0]$ \\
\bottomrule
\end{tabular}
\end{table}

\begin{table*}[t]
\centering
\caption{Downstream performance under linear probing. Both LatentWave variants are pretrained with $N_m = 6$ target masks. Classification tasks report mean per-class accuracy (\%, $\uparrow$); positioning reports mean positioning error (m, $\downarrow$). Results are mean $\pm$ std over three seeds. Best self-supervised result in \textbf{bold}.}
\label{tab:main_results}
\resizebox{\textwidth}{!}{%
\begin{tabular}{l c c c | c}
\toprule
\textbf{Task} &
\textbf{\shortstack{LatentWave\\(Region)}} &
\textbf{\shortstack{LatentWave\\(Freq.)}} &
\textbf{WavesFM} &
\textbf{Supervised} \\
\midrule
RF Signal Classification &
$\mathbf{80.9 \pm 2.94}$ & $66.1 \pm 3.10$ & $80.3 \pm 0.73$ &
$86.5 \pm 1.69$ \\
5G NR Positioning (m) &
$2.54 \pm 0.009$ & $\mathbf{2.32 \pm 0.023}$ & $2.77 \pm 0.037$ &
$0.71 \pm 0.001$ \\
Beam Prediction &
$51.6 \pm 0.35$ & $\mathbf{63.1 \pm 0.45}$ & $51.2 \pm 0.36$ &
$88.9 \pm 0.50$ \\
LoS/NLoS Classification &
$92.9 \pm 0.57$ & $\mathbf{93.4 \pm 0.26}$ & $93.4 \pm 0.54$ &
$95.9 \pm 0.27$ \\
\bottomrule
\end{tabular}%
}
\end{table*}

\begin{figure}
    \centering
    \includegraphics[width=\linewidth]{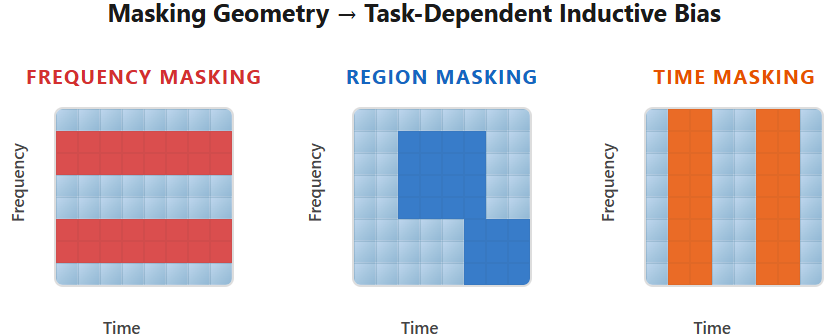}
    \caption{ The three structured JEPA mask geometries (region, frequency, and time) each inducing a different wireless inductive bias.}
    \label{fig:JepaMasks}
\end{figure}

\subsection{Masking Strategy}
\label{subsec:masking}

The masking strategy defines the prediction problem faced during pretraining: what information is hidden, what context remains, and which structures the model must infer. We investigate four strategies, each imposing a distinct inductive bias on the learned representation. Fig.~\ref{fig:JepaMasks} illustrates the three structured geometries: region, frequency, and time masking.
In all variants, the context visible to the encoder is obtained by removing the target masks and sampling a context block of scale $s_c \in [s_{c,\min}, s_{c,\max}]$ from the remaining patches. In all cases, masking a patch at a given time-frequency position removes the corresponding tokens across all sampled channels, ensuring a consistent prediction target in the spatial domain.

\textbf{Region masking.} Following I-JEPA~\cite{jepa}, $N_m$ two-dimensional masks are generated by sampling a target scale $s \in [s_{\min}, s_{\max}]$ and an aspect ratio $a \in [a_{\min}, a_{\max}]$, which together determine the mask dimensions along the frequency and time axes. Each mask is placed at a uniformly random position within the spectrogram. Masks may overlap with one another but not with the context region. This encourages the model to capture joint spectro-temporal structure, such as modulation patterns and bandwidth occupancy.

\textbf{Frequency masking.} Each of the $N_m$ masks spans the \emph{entire time axis} while covering a contiguous subset of frequency bins. The model must reconstruct complete spectral bands from the remaining frequency content, imposing a bias toward learning inter-frequency dependencies and spectral envelope structure.

\textbf{Time masking.} Each of the $N_m$ masks spans the \emph{entire frequency axis} while covering a contiguous subset of time steps. The model must predict full spectral snapshots at hidden time instants, encouraging the capture of temporal dynamics such as burst timing and temporal correlations across frames.

\textbf{Random masking.} As in WavesFM~\cite{wavesfm}, individual patches are sampled uniformly at random without any spatial contiguity constraint. This provides maximal spatial diversity but allows local interpolation from neighboring patches to partially solve the prediction task.

For the region, frequency, and time strategies, we vary $N_m \in \{2, 3, 4, 5, 6\}$ to control the total masked area. A systematic comparison of all strategies and mask counts is presented in Section~\ref{subsec:ablation}.

\subsection{Pretraining Data}

We pretrain on four unlabeled datasets spanning spectrograms and CSI. The first is a privately collected spectrogram dataset that contains 3200 WiFi, LTE, Bluetooth, 5G NR, and ISM signals captured over the air using \ac{sdr} at various sub-6\,GHz center frequencies with sampling rates of 10--60\,MHz. The second is an RF fingerprinting spectrogram dataset comprising approximately 6000 single-channel samples~\cite{powder}. The third is a 5G NR indoor CSI dataset with approximately 15000 five-channel samples~\cite{effecientfi}. The fourth is a WiFi CSI dataset with 840 three-channel samples~\cite{wifi-csi}. Together, the pretraining corpus totals roughly 25000 samples with varying numbers of channels. All spectrogram samples are converted to the decibel scale as $20\log_{10}(\cdot)$. Every sample, whether spectrogram or CSI, is then resized to $224 \times 224$ and normalized using per-channel mean and standard deviation computed over each pretraining dataset.

\subsection{Baseline and Downstream Evaluation}

We use WavesFM~\cite{wavesfm} as the primary baseline, since it was pretrained with masked modeling. We reproduced it on the same pretraining datasets and evaluated on the same downstream tasks. This controlled comparison isolates the effect of the \ac{jepa} pretraining objective on representation quality.

We evaluate on four downstream tasks: RF signal classification using CommRad~RF~\cite{commrad}, 5G NR positioning using 5G~CSI (the outdoor scenario, distinct from the indoor scenario used during pretraining), beam prediction, and LoS/NLoS classification both using DeepMIMO~\cite{deepmimo}. These tasks span communication, sensing, and positioning domains, with several distributions and modalities that differ from the pretraining data. For each task, we evaluate using linear probing where the encoder is frozen, and a single linear head is trained on the mean-pooled patch representations for LatentWave and the CLS token for WavesFM. 

\begin{table*}[t]
\centering
\caption{Masking strategy ablation (one seed). The effective masking ratio for each configuration is shown in the third header row. Evaluation metrics are identical to Table~\ref{tab:main_results}. Best result per task in \textbf{bold}.}
\label{tab:ablation}
\renewcommand{\arraystretch}{1.15}
\setlength{\tabcolsep}{4pt}
\small
\resizebox{\textwidth}{!}{%
\begin{tabular}{l ccccc ccccc ccccc c}
\toprule
& \multicolumn{5}{c}{\textbf{Region}} &
  \multicolumn{5}{c}{\textbf{Frequency}} &
  \multicolumn{5}{c}{\textbf{Time}} &
  \textbf{Rand.} \\
\cmidrule(lr){2-6} \cmidrule(lr){7-11} \cmidrule(lr){12-16} \cmidrule(lr){17-17}
$N_m$ &
2 & 3 & 4 & 5 & 6 &
2 & 3 & 4 & 5 & 6 &
2 & 3 & 4 & 5 & 6 & -- \\
Mask ratio &
\multicolumn{1}{c}{\scriptsize 29\%} & \multicolumn{1}{c}{\scriptsize 43\%} &
\multicolumn{1}{c}{\scriptsize 57\%} & \multicolumn{1}{c}{\scriptsize 71\%} &
\multicolumn{1}{c}{\scriptsize 80\%} &
\multicolumn{1}{c}{\scriptsize 29\%} & \multicolumn{1}{c}{\scriptsize 43\%} &
\multicolumn{1}{c}{\scriptsize 57\%} & \multicolumn{1}{c}{\scriptsize 71\%} &
\multicolumn{1}{c}{\scriptsize 80\%} &
\multicolumn{1}{c}{\scriptsize 29\%} & \multicolumn{1}{c}{\scriptsize 43\%} &
\multicolumn{1}{c}{\scriptsize 57\%} & \multicolumn{1}{c}{\scriptsize 71\%} &
\multicolumn{1}{c}{\scriptsize 80\%} &
\multicolumn{1}{c}{\scriptsize 75\%} \\
\midrule
RF Signal Classif. &
73.3 & 73.1 & 74.4 & 79.7 & \textbf{83.5} &
74.6 & 73.9 & 68.9 & 68.0 & 68.3 &
76.0 & 78.2 & 77.7 & 80.2 & 81.5 & 71.2 \\
5G NR Positioning (m) &
3.19 & 2.71 & 2.30 & 2.45 & 2.55 &
3.20 & 2.95 & 2.42 & 2.36 & \textbf{2.32} &
3.43 & 3.54 & 3.43 & 3.18 & 2.84 & 2.90 \\
Beam Prediction &
37.2 & 40.7 & 46.5 & 48.6 & 51.6 &
44.6 & 45.6 & 49.8 & 59.5 & \textbf{63.5} &
33.6 & 38.5 & 37.4 & 38.7 & 44.6 & 39.2 \\
\bottomrule
\end{tabular}
}
\end{table*}

\section{Results}
We now present and discuss the results. 
Section~\ref{subsec:setup} details the evaluation metrics and implementation specifics. Section~\ref{subsec:main_results} compares LatentWave against supervised and self-supervised baselines under linear probing. Section~\ref{subsec:ablation} ablates the masking strategies to justify the design choices.

\subsection{Experimental Setup}
\label{subsec:setup}

For all downstream tasks, we split the data into 80\% training and 20\% test sets. We report mean per-class accuracy for all classification tasks, namely RF signal classification, beam prediction, and LoS/NLoS classification, as it accounts for class imbalance more faithfully than overall accuracy. Given \(K\) classes, the mean per-class accuracy is defined as
\begin{equation}
    \mathrm{Acc}_{\mathrm{mpc}} = \frac{1}{K} \sum_{k=1}^{K} \frac{n_k^{\mathrm{correct}}}{n_k},
\end{equation}
where \(n_k\) is the total number of test samples belonging to class \(k\) and \(n_k^{\mathrm{correct}}\) is the number of correctly classified samples in that class. For the positioning task, we report the mean positioning error, defined as the average Euclidean distance between the predicted and ground-truth coordinates:
\begin{equation}
    \mathrm{MPE} = \frac{1}{N} \sum_{i=1}^{N} \bigl\| \hat{\mathbf{p}}_i - \mathbf{p}_i \bigr\|_2,
\end{equation}
where \(N\) is the number of test samples, \(\hat{\mathbf{p}}_i \in \mathbb{R}^3\) is the predicted position, and \(\mathbf{p}_i \in \mathbb{R}^3\) is the ground-truth position of the \(i\)-th sample.

For all tasks, the frozen encoder embeddings are first standardized to zero mean and unit variance using statistics computed on the training split. A logistic regression classifier is used for classification tasks, and linear regression for positioning. Main results are reported as the mean and standard deviation over three random seeds, controlling the train/test split.

\subsection{Main Results} 
\label{subsec:main_results}

Table~\ref{tab:main_results} compares the downstream performance of all methods under linear probing. LatentWave with region masking performs on par with WavesFM across all tasks, with marginal gains on RF signal classification and positioning, indicating that predicting in latent space matches pixel-level reconstruction in representation quality.

Switching to frequency masking yields a markedly different profile: beam prediction improves by over 11 percentage points and positioning error decreases from 2.54\,m to 2.32\,m, but RF signal classification drops substantially from 80.9\% to 66.1\%. This suggests that frequency masking encourages representations that capture spatial and spectral structure beneficial for channel-related tasks, at the cost of temporal patterns needed to discriminate signal types. LoS/NLoS classification remains largely unaffected across all strategies.

Despite using only a frozen encoder with a linear head, the self-supervised methods closely approach the supervised upper bound on RF classification and LoS/NLoS classification, with the remaining gap on beam prediction and positioning suggesting clear benefit from scaling pretraining data or exploring lightweight fine-tuning.

\subsection{Masking Ablations}
\label{subsec:ablation}

Table~\ref{tab:ablation} ablates the masking strategy and the number of masks $N_m$ across three downstream tasks. The effective masking ratio, i.e., the average proportion of patches hidden from the context encoder, increases with $N_m$ from 29\% at $N_m{=}2$ to 80\% at $N_m{=}6$.

Increasing $N_m$ consistently improves beam prediction across all structured strategies. RF signal classification follows the opposite trend under frequency masking, degrading as more spectral bands are hidden, whereas region and time masking both improve with more masks. This confirms that frequency masking biases representations toward inter-frequency structure at the expense of temporal discriminability needed for signal type recognition. Positioning error generally decreases with more masks for frequency and time strategies, while region masking achieves its best positioning result at $N_m{=}4$.

Random masking at 75\% performs competitively with low-mask-count structured strategies but is consistently outperformed by higher mask counts, confirming that contiguous masking with sufficient coverage yields stronger representations. Overall, no single strategy dominates all tasks.

\section{Conclusion}
This paper presented LatentWave, a JEPA-based wireless foundation model that learns by predicting masked regions in latent space. The architecture decouples the encoder from a fixed antenna count through per-channel patch embedding and stochastic channel sampling, enabling a single pretrained model to handle both spectrogram and CSI inputs with varying numbers of channels. We also showed that the masking geometry introduces a task-dependent inductive bias: frequency masking favors channel-related tasks such as positioning and beam prediction, while region masking better preserves discriminability for signal classification. These findings position masking strategy design as a promising direction for building more transferable wireless foundation models.

\balance

\bibliography{bibliography.bib}
\bibliographystyle{ieeetr}

\end{document}